\documentclass[twocolumn,aps,prl,showpacs]{revtex4}
\usepackage{graphicx}
\DeclareGraphicsExtensions{ps,eps,epsf}

\begin{document}

\title{Momentum-resolved electron-phonon interaction in lead determined by neutron resonance spin-echo spectroscopy}

\author{T. Keller$^{1,2}$, P. Aynajian$^1$, K. Habicht$^3$,
L. Boeri$^1$, S.K. Bose$^4$, and B. Keimer$^1$ }

\affiliation{$^1$ Max-Planck-Institut f\"ur Festk\"orperforschung,
Heisenbergstr. 1, D-70569 Stuttgart, Germany}

\affiliation{$^2$ ZWE FRMII, Technical University of Munich,
Lichtenbergstr. 1, D-85748 Garching, Germany}

\affiliation{$^3$ Hahn-Meitner-Institut, Glienicker Str. 100,
D-14109 Berlin, Germany}

\affiliation{$^4$ Brock University, 500 Glenridge Avenue, St.
Catharines, Ontario L2S 3A1, Canada}

\date{\today}

\begin{abstract}
Neutron resonance spin-echo spectroscopy was used to monitor the
temperature evolution of the linewidths of transverse acoustic
phonons in lead across the superconducting transition temperature,
$T_c$, over an extended range of the Brillouin zone. For phonons
with energies below the superconducting energy gap, a linewidth
reduction of maximum amplitude $\sim 6 \mu$eV was observed below
$T_c$. The electron-phonon contribution to the phonon lifetime
extracted from these data is in satisfactory overall agreement
with {\it ab-initio} lattice-dynamical calculations, but
significant deviations are found.
\end{abstract}


\maketitle

The electron-phonon interaction is the major limiting factor for
electronic transport phenomena in metals at elevated temperatures,
and it is responsible for most instances of superconductivity. For
metallic and superconducting elements \cite{savrasov} and binary
compounds such as MgB$_2$ \cite{mgb2}, modern {\it ab initio}
calculations yield accurate predictions for the electron-phonon
coupling parameters of every phonon over the entire Brillouin zone
(BZ). An active area of research is aimed at a realistic description
of electron-lattice interactions in complex compounds with strong
electronic correlations. Yet only rudimentary experimental tests of
these calculations have thus far proven possible. For instance,
electronic transport or tunnelling experiments probe weighted
averages of the electron-phonon interaction over the entire phonon
spectrum. Optical spectroscopy is capable of probing the lifetimes
of individual phonons limited by scattering from electrons, but
kinematics constrains these experiments to a single point in
momentum space. Here we use a new neutron spectroscopy method
\cite{golub,keller1,bayrakci} with an energy resolution in the
$\mu$eV range (that is, about two orders of magnitude better than
that of standard neutron spectroscopy) to determine the
electron-phonon lifetime of an individual acoustic phonon in lead
over an extended range of the BZ. The results are compared to {\it
ab initio} lattice dynamical calculations also reported here. Our
experiment constitutes the first detailed test of modern
calculations of the electron-phonon interaction and opens up a new
avenue for a quantitative understanding of electron-phonon
interactions and superconductivity in solids.

Since the 1950's, triple-axis spectrometry (TAS) with neutrons
(and recently, x-rays) has been the method of choice to
experimentally determine energy- and momentum-resolved phonon
spectra of solids. Briefly, the energies and momenta of the
incoming and scattered neutrons are selected by crystal
monochromators, and their difference yields the phonon dispersion
relation. The monochromaticity of the beam is thus coupled to the
beam divergence, which can only be restricted at the expense of
beam intensity. This implies that energy resolutions significantly
better than 10\% are impractical to achieve under almost all
circumstances. Since the electron-phonon interaction leads to a
phonon linewidth broadening of typically less than 1\%, TAS allows
phonon lifetime measurements only in exceptional cases. While the
phonon dispersions in lead were determined by TAS early on
\cite{brockhouse}, attempts to resolve the electron-phonon
lifetimes have been unsuccessful \cite{furrer,youngblood}. TAS is
therefore not generally applicable as a probe of the
electron-phonon interaction in metals.

A neutron spin-echo method to improve the energy resolution of TAS
by several orders of magnitude without loss of intensity was
proposed some time ago \cite{nse}. The method has been described
in detail elsewhere \cite{keller1}. Briefly, it is implemented by
inserting spin polarizers and tunable Larmor precession coils into
the incident and scattered beams of the triple-axis spectrometer.
If the phonon to be measured has infinite lifetime, and if the
precession fields are properly matched (spin-echo condition), the
full polarization of the incident beam is recovered at the
detector. If the phonon lifetime is finite, the spin-echo
amplitude is reduced, and the phonon linewidth can be determined
by systematically varying the precession fields. Early versions of
this technique have enabled the determination of the lifetimes of
rotons in superfluid helium \cite{mezei} and optical phonons near
the BZ boundary in germanium \cite{kulda}. However, a major
technical obstacle has thus far prevented more general
applications. The dispersive nature of most collective excitations
in solids leads to a degradation of the neutron spin polarization
that masks the effect of the finite excitation lifetime. This
effect can be compensated to linear order by tilting the
precession coils away from the neutron beam direction by an angle
proportional to the slope of the dispersion relation
\cite{keller1}. Since the required tilt angles are of order
10-50$^\circ$, however, this cannot be accomplished by the long
solenoids used in the early work.

In the ``neutron resonance spin-echo" (NRSE) technique
\cite{golub,keller1,bayrakci}, the solenoids are replaced by a pair
of compact radio-frequency ({\it rf}$\,$) coils surrounding a
field-free region. This allows larger tilt angles sufficient to
match the dispersion relations of most collective excitations in
solids. Using a prototype NRSE-TAS setup, the feasibility of phonon
lifetime measurements in lead was demonstrated at temperatures
exceeding 50 K, where the lifetime is limited by phonon
anharmonicity \cite{habicht1}. However, the neutron flux was
insufficient to obtain high-quality data at lower temperatures,
where the electron-phonon interaction is dominant.

Here we report the results of experiments carried out at the
dedicated, high-flux NRSE-TAS spectrometer TRISP at the FRM-II
neutron source in Garching, Germany \cite{keller2,bayrakci}.
Starting from the source, in-beam components include a
spin-polarizing guide; a velocity selector to cut out higher-order
contamination of the incident beam; a horizontally and vertically
focusing pyrolytic-graphite (PG) monochromator and a horizontally
focusing PG analyzer, each set for the (002) reflection; and a
supermirror spin-polarizer in front of the detector. The
measurements closely followed the protocol established in Ref.
\onlinecite{habicht1}. The reader is advised to consult this
reference for a detailed description of the experimental setup and
data analysis. The samples were two cylindrical Pb single crystals
with cylinder axes (100) and (110) and mosaicities 4.2' and 5.7',
respectively. Both crystals had a diameter of 30 mm and a length of
50 mm. The first crystal was used to measure the $T_1$-phonon along
(110) and the $T$-phonon along (100) in the scattering plane spanned
by the (001) and (010) directions of the reciprocal lattice. The
remaining measurements were carried out with the second sample. The
samples were loaded into a closed-cycle He cryostat. Depending on
phonon energy, the incident (final) neutron wave vector was fixed at
$\rm 2.51 \AA ^{-1}$ ($\rm 1.7 \AA ^{-1}$). The triple-axis
spectrometer was aligned in a defocusing configuration in order to
minimize the segment of the dispersion surface sampled by the TAS
resolution ellipsoid. The spectrometer angles were tuned to maximize
the phonon intensity and kept constant during the spin-echo scans.
The NRSE coils between monochromator and sample and between sample
and analyzer were tilted to match the slope of the acoustic phonon
dispersions. The spin-echo time, $\tau$, was changed by varying the
{\it rf}$\,$ frequencies in the NRSE coils. In order to minimize
systematic errors, the beam polarization at the detector, $P$, was
determined by translating the coil immediately in front of the
detector and extracting the amplitude of the resulting sinusoidal
intensity modulation from a least-squares fit \cite{habicht1}.

The strategy of the experiment was to single out the
electron-phonon contribution to the phonon lifetime by monitoring
the evolution of the phonon lineshape through the superconducting
transition temperature, $T_c = 7.2$ K. As demonstrated previously
by TAS for phonons with anomalously large linewidths in Nb$_3$Sn
\cite{axe} and Nb \cite{shapiro}, this contribution vanishes if
the phonon energy is below the superconducting energy gap, $2
\Delta$. Other intrinsic contributions to the phonon linewidth,
such as isotope disorder and the anharmonicity of the lattice
potential, do not exhibit anomalies at $T_c$.

Because of its face-centered cubic lattice structure, the phonon
spectrum of lead consists entirely of acoustic branches. Most of
the data were taken on the lowest-lying transverse mode, $T_1$, in
the $(\xi \xi 0)$ direction, where all phonons are nondegenerate.
The atomic displacements of this mode are along
$(1\overline{1}0)$. The mode energy is lower than $2 \Delta$ over
much of the BZ, so that a large range of phonon momenta can be
covered in this way. More limited data sets were also acquired for
the $T_2$ mode (with displacements in the (001) direction) along
$(\xi \xi 0)$, and for the transverse phonon along $(\xi \xi
\xi)$.

\begin{figure}[t]
\includegraphics[width=9.0cm]{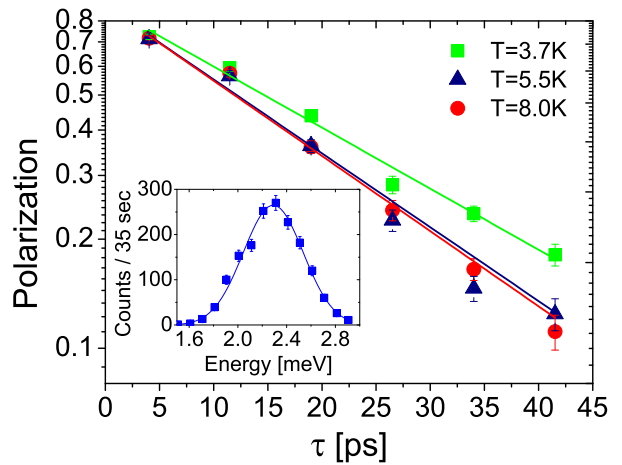}
\caption{(Color online) Neutron resonance spin echo profiles for
the $T_1$-phonon at $(0.3, 0.3, 0)$ at selected temperatures above
and below the superconducting transition temperature. The inset
shows a triple-axis scan in focusing mode for the same phonon.
From these data, $\Gamma$ was extracted as $25.5 \pm 0.6$, $31.4
\pm 0.8$, and $30.9 \pm 0.7$ $\mu$eV for $T=3.7$, 5.5, and 8.0 K,
respectively.
 } \label{fig1}
\end{figure}

\begin{figure}[t]
\includegraphics[width=9.0cm]{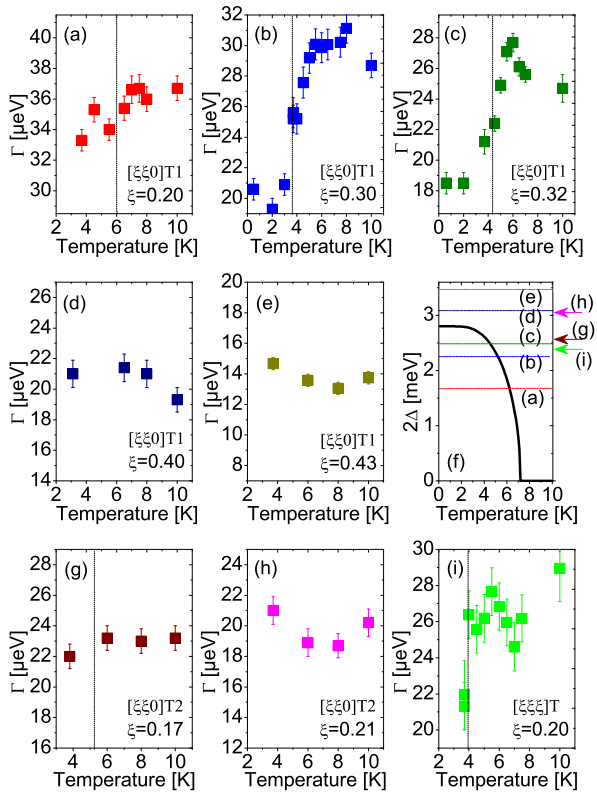}
\caption{(Color online) a-e) Temperature dependence of the
observed NRSE relaxation rate $\Gamma$ for $T_1$-phonons along
$(\xi \xi 0)$. f) Superconducting energy gap from tunnelling data
\cite{gasparovic}. g,h) $\Gamma$ versus temperature for
$T_2$-phonons along $(\xi \xi 0)$, and i) for a transverse phonon
along $(\xi \xi \xi)$. The dashed lines in panels a-e, g, i
indicate the temperatures at which the superconducting gap opens,
according to the data of panel f. } \label{fig2}
\end{figure}

Fig. 1 shows typical NRSE data for the $T_1$ mode along $(\xi \xi
0)$. The dependence of the beam polarization on the spin-echo
time, $P(\tau)$, is proportional to the Fourier transform of the
phonon lineshape \cite{keller1}. As a Lorentzian lineshape thus
yields an exponential NRSE profile, the phonon linewidth can be
obtained from the slope, $\Gamma$, of $(P(\tau)$ on a
semilogarithmic plot (lines in Fig. 1). A drop of the slope
(corresponding to an increase of the phonon lifetime) below $T_c$
is clearly apparent. In addition to the intrinsic phonon
linewidth, $\Gamma$ is also affected by extrinsic factors such as
the mosaicity of the crystal and the curvature of the dispersion
relation, as discussed in detail in Refs. \cite{keller2,habicht2}.
However, these temperature-independent factors have a negligible
influence on the linewidth anomaly at $T_c$, which is of interest
here. A detailed discussion of resolution effects on the present
data will thus be given elsewhere \cite{note}. In the following
figures presenting the experimentally observed $\Gamma$ versus
temperature, the extraneous contributions give rise to
zero-temperature offsets that generally decrease with increasing
wave vector, due to the decreasing curvature of the dispersion
surface sampled by the TAS resolution volume.

Fig. 2a-e shows the temperature dependence of $\Gamma$ for
selected $T_1$-phonons along $(\xi \xi 0)$. A decrease of $\Gamma$
at $T_c$ is evident in all of the traces shown for $\xi \leq 0.32$
($E \leq 2.46$ meV). The amplitude of the effect increases with
increasing $\xi$ and reaches $\sim 6 \mu$eV for $\xi = 0.32$. For
$\xi = 0.4$ ($E = 3.05$ meV), the effect is no longer present
within the experimental error. This observation is in good accord
with the magnitude of the superconducting energy gap, $2 \Delta
\sim 2.6$ meV, extracted from tunnelling data \cite{gasparovic}.
For $\xi = 0.43$, there is some indication of an increase in
phonon linewidth at low temperatures. This is expected because of
the pileup of electronic states above the energy gap. This effect
is also responsible for the nonmonotonic temperature dependence of
$\Gamma$ for $\xi = 0.32$ (Fig. 2c). As $2\Delta$ approaches the
phonon energy $E$ from below with decreasing temperature, $\Gamma$
first increases gradually because of the enhanced density of
states above the gap, and subsequently plummets sharply below
$T_c$ as $2\Delta > E$. It has been shown \cite{shapiro,bobetic}
that the jump in phonon linewidth at $T_c$, expressed as a
fraction of the normal-state linewidth, is equal to the jump in
the ultrasound attenuation coefficient at $T_c$. This in turn can
be calculated in BCS theory \cite{bobetic} as $(\pi/2)[1-2/(1+
\exp(E/k_B T))] = 1.44$ for $E=2.46$ meV and $T=4.6$ K (the
temperature where $2\Delta = E$), in good agreement with the value
$1.48 \pm 0.1$ determined from Fig. 2c.

For the more steeply dispersing $T_2$ phonon along $(\xi \xi 0)$,
the superconductivity-induced phonon linewidth renormalization is
much smaller than that of the $T_1$ phonon, and only upper bounds
could be established (Fig. 2g). This is expected on general
grounds, because the steep dispersion of this mode implies that
the linewidth effects are restricted to wave vectors close to the
BZ center. In the $(\xi \xi \xi)$ direction, where the two
transverse phonons are degenerate, we observed a
superconductivity-induced linewidth reduction of magnitude similar
to that of the $T_1$-phonon along $(\xi \xi 0)$ (Fig. 2i).

In order to compare the experimental results to predictions of
{\it ab-initio} calculations, the phonon dispersions and
linewidths were calculated in the framework of density functional
\cite{DFT:KS,DFT:HK} perturbation theory \cite{DFT:BaroniRMP} in
the local-density approximation. We employed ultra-soft
pseudopotentials~\cite{Vanderbilt}, with scalar-relativistic
corrections and a plane-wave basis set~\cite{PWscf}, with a
cut-off energy of 40 Ryd. The {\bf k}-space integration was
approximated with a $(12)^3$ Monkhorst-Pack grid~\cite{MPgrid}
using a smearing parameter ~\cite{Coldsmearing} of 0.06 Ryd for
the self-consistent cycles and phonon calculations, and with a
much denser $(36)^3$ mesh and a Gaussian smearing of 0.04 Ryd for
the calculation of the linewidths. With these parameters and the
numerically optimized lattice parameter of $\rm 4.95 \AA$, the
phonon frequencies were converged to 0.1 meV and the linewidths to
0.025 $\mu$eV. (If the lattice parameter $\rm 4.92 \AA$ determined
experimentally at low temperature is used instead, the results are
not affected outside these numerical confidence limits.) The
calculated phonon frequencies agree with the experimentally
determined phonon frequencies \cite{brockhouse,furrer} within the
published error bars. For selected $\mathbf{k}$-points, we
recalculated the phonon energies and linewidths using Savrasov's
full potential LMTO program \cite{savrasov} and found very good
agreement.

%\begin{figure}[t]
%\includegraphics[width=9.0cm]{fig3.EPS}
%\caption{(Color online) Temperature dependence of the observed
%NRSE relaxation rate $\Gamma$ for a) $T_2$-phonons along $(\xi \xi
%0)$, and b) a transverse phonon along $(\xi \xi \xi)$. The dashed
%lines indicate the temperatures at which the superconducting gap
%opens (see Fig. 2f).
% } \label{fig3}
%\end{figure}

\begin{figure}[t]
\includegraphics[width=9.0cm]{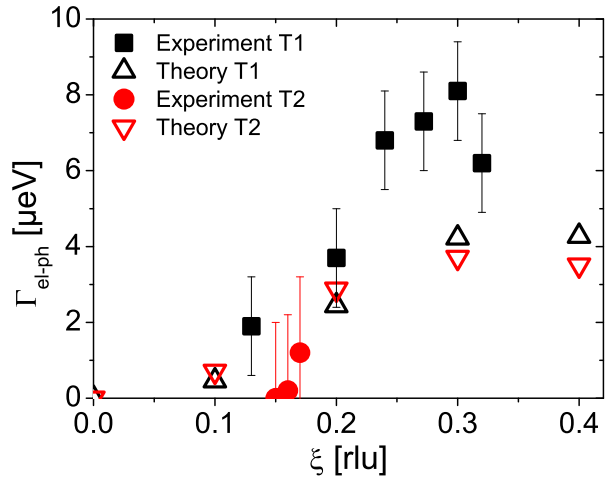}
\caption{(Color online) Experimentally determined and numerically
calculated electron-phonon linewidth of transverse phonons along
$(\xi \xi 0)$. The experimental points were obtained by
subtracting the linewidths at 10K (see Fig. 2) from those at the
lowest temperature (0.5 K).
 } \label{fig3}
\end{figure}

%The phonon dispersions  were obtained by Fourier interpolation of
%the dynamical matrices calculated on a
% a $(12)^3$ grid in reciprocal space.
%The total electron-phonon coupling parameter, estimated by summing
%over all phonon branches and wavevector, is $1.22$, which is lower
%than the experimental value of $1.55$\cite{Savrasov}, but in good
%agreement with previous pseudopotential calculations\cite{Liu}.
%The inclusion of spin-orbit coupling should improve the agreement
%between the calculated and the measured value.

Fig. 4 provides a synopsis of the experimentally determined and
numerically calculated electron-phonon linewidths along $(\xi \xi
0)$. Although the overall agreement is satisfactory, a significant,
systematic deviation outside of the experimental error bar is
apparent around $\xi=0.3$. A similar deviation was found for the
more limited data set along $(\xi \xi \xi)$. A possible source of
this discrepancy is the spin-orbit coupling, which was not
considered in the calculation. The electronic density of states at
the Fermi level increases by about 10\% when spin-orbit coupling is
incorporated in the electronic structure calculation. A
corresponding increase in the electron-phonon interaction is to be
expected. Indeed, Pb is a well-known example in which the agreement
between the experimentally determined and theoretically calculated
phonon frequencies and total electron-phonon coupling parameter
$\lambda$ is not as good as in the case of other elemental
superconductors, such as Al or Nb. In our calculation, $\lambda$ was
computed as 1.27, whereas $\lambda = 1.55$ was determined by
tunneling spectroscopy \cite{gasparovic}. Furthermore, the values of
$\lambda$ calculated from first principles range from $\lambda=1.2$
(Ref. \onlinecite{liu}) to $\lambda=1.68$ (Ref.
\onlinecite{savrasov}). The experimental results presented here
therefore call for further development in the numerical computation
of electron-phonon interactions.

In conclusion, we have used neutron resonance spin-echo spectroscopy
\cite{golub,keller1,bayrakci} to determine the electron-phonon
contribution to the lifetimes of selected phonons in lead. The
results have enabled the first energy- and momentum-resolved test of
{\it ab initio} lattice-dynamical calculations. The energy
resolution of the experiment was about two orders of magnitude
better than that achievable by triple-axis spectroscopy (for a
typical triple-axis scan, see the inset in Fig. 1). The method can
be applied in a straightforward manner to other elements and
compounds that exhibit superconductivity at low temperatures. For
non-superconducting metals, a rigorous treatment of the resolution
function is required to extract the electron-phonon lifetime. The
pertinent formalism has already been established
\cite{habicht1,habicht2}. Finally, the technique has recently been
successfully applied to magnetic excitations \cite{bayrakci} and
opens many new perspectives in this field as well.

We would like to thank O.K. Andersen, S.P. Bayrakci, and F. Mezei
for useful discussions, and C.T. Lin for preparing the Pb crystals.

\bibliographystyle{prsty}

\begin{thebibliography}{10}

\bibitem{savrasov} See, {\it e.g.}, S. Y. Savrasov, Phys. Rev. B
\textbf{54}, 16470 (1996); S.Y. Savrasov and D.Y. Savrasov, {\it
ibid.} {\bf 54}, 16487 (1996);  S.Y. Savrasov, D.Y. Savrasov, and
O.K. Andersen, Phys. Rev. Lett. {\bf 72}, 372 (1994).

\bibitem{mgb2} See, {\it e.g.}, H.J. Choi, D. Roundy, H. Sun, M.L.
Cohen, and S.G. Louie, Nature {\bf 418}, 758 (2002); Y. Kong, O.V.
Dolgov, O. Jepsen, and O.K. Andersen, Phys. Rev. B {\bf 64}, 020501
(2001).

\bibitem{golub} R. Golub and R. G\"ahler, Phys. Lett. A {\bf 123}, 43 (1987).

\bibitem{keller1} T. Keller, B. Keimer, K. Habicht, R. Golub, and F. Mezei in:
{\it Neutron Spin Echo Spectroscopy} (Springer Lecture Notes in
Physics 601), edited by F. Mezei, C. Pappas, and T. Gutberlet
(Springer, Heidelberg, 2003), p. 74. Available online at
http://www.fkf.mpg.de/keimer/.

\bibitem{bayrakci} S.P. Bayrakci, T. Keller, K. Habicht, and B.
Keimer, to be published.


\bibitem{brockhouse} B.N. Brockhouse, Phys. Rev. {\bf 128}, 1099
(1962); R. Stedman, L. Almquist, G. Nilsson, and G. Raunio, {\it
ibid.} {\bf 162}, 161 (1967).

\bibitem{furrer} A. Furrer and W. H\"alg, Phys. Stat. Sol. {\bf
42}, 821 (1970).

\bibitem{youngblood} R. Youngblood, Y. Noda, and G. Shirane,
Sol. State Commun. {\bf 21}, 1433 (1978).



\bibitem{nse} F. Mezei, {\it Inelastic Neutron Scattering} (IAEA, Vienna, 1978), p.
125; R. Pynn, J. Phys. E {\bf 11}, 1133 (1978).

\bibitem{mezei} F. Mezei, B. Farago, and C. Lartigue, in {\it Excitations in Two-
Dimensional and Three-Dimensional Quantum Fluids}, Vol. 257 of
{\it NATO Advanced Studies Institute, Series B: Physics}, edited
by A.F.G. Wyatt and H.J. Lauter (Plenum, New York, 1991), p. 119.

\bibitem{kulda} J. Kulda, A. Debernardi, M. Cardona, F. de Geuser, and E.E.
Haller, Phys. Rev. B {\bf 69}, 045209 (2004).



\bibitem{habicht1} K. Habicht, R. Golub, F. Mezei, B. Keimer, and T.
Keller, Phys. Rev. B {\bf 69}, 104301 (2004).

\bibitem{keller2} T. Keller, K. Habicht, H. Klann, M. Ohl, H. Schneider, and B.
Keimer, Appl. Phys. A {\bf 74}, S332 (2002).

\bibitem{habicht2} K. Habicht, T. Keller, and R. Golub, J. Appl. Crystallogr. {\bf 36},
1307 (2003).

\bibitem{axe} J.D. Axe and G. Shirane, Phys. Rev. Lett. {\bf 30},
214 (1972); Phys. Rev. B {\bf 8}, 1965 (1973).

\bibitem{shapiro} S.M. Shapiro, G. Shirane, and J.D. Axe, Phys.
Rev. B {\bf 12}, 4899 (1975).

\bibitem{note} P. Aynajian {\it et al.}, to be published.

\bibitem{gasparovic} R.F. Gasparovic, B.N. Taylor, and R.K. Eck,
Sol. State Commun. {\bf 4}, 59 (1966).

\bibitem{bobetic} V. M. Bobetic, Phys. Rev. {\bf 136}, A1535
(1964).

\bibitem{DFT:KS}
W.~Kohn and L.~J. Sham, Phys. Rev. \textbf{40}, A1133 (1965).

\bibitem{DFT:HK}
P.~Hohenberg and W.~Kohn, Phys. Rev. \textbf{136}, B864 (1964).

\bibitem{DFT:BaroniRMP}
S.~Baroni, S.~de~Gironcoli, A.~Dal Corso, and P.~Giannozzi, Rev.
Mod. Phys. \textbf{73}, 515 (2001).

\bibitem{Vanderbilt}
D.~Vanderbilt, Phys. Rev. B \textbf{41}, 7892 (1990).

\bibitem{PWscf}
S. Baroni {\it et al.}, http://www.pwscf.org.

\bibitem{MPgrid}
H.J. Monkhorst and J.D. Pack, Phys. Rev. B \textbf{13}, 5188
(1976).

\bibitem{Coldsmearing}
N. Marzari, D. Vanderbilt, A. De Vita, and M.C. Payne, Phys. Rev.
Lett. \textbf{82}, 3296 (1999).

\bibitem{liu} A.Y. Liu and A.A. Quong, Phys. Rev. B {\bf 53}, R7575
(1996).







\end{thebibliography}

\end{document}